\documentclass[12pt,a4]{article}
\pdfoutput=1
\usepackage{jheppub}
\usepackage{latexsym,diagbox,stackrel}
\usepackage{mathrsfs}
\usepackage{bbold}
\usepackage{pdflscape}
\usepackage{amssymb}
\usepackage{amsmath}
\usepackage{multirow}
\usepackage{cleveref}
\usepackage{tikz}
\usepackage{graphicx}
\usetikzlibrary{shapes.geometric, arrows}

\usepackage
[
color=gray]
{todonotes} 

\newcommand{\half}{\frac{\scriptstyle 1}{\scriptstyle 2}}

\newcommand{\C}{\mathbb{C}}

\newcommand{\CP}{\mathbb{CP}}

\newcommand{\p}{\partial}
\newcommand{\dbar}{\bar\partial}
\newcommand{\e}{\mathrm{e}}
\newcommand{\g}{\mathfrak{g}}

\newcommand{\cG}{\mathcal{G}}

\newcommand{\cC}{\mathcal{C}}

\newcommand{\cM}{\mathcal{M}}

\newcommand{\cX}{\mathcal{X}}

\newcommand{\tr}{\mathrm{tr}}

\newcommand{\SU}{\, \mathrm{SU}}

\newcommand{\rd}{\, \mathrm{d}}
\newcommand{\Pf}{\, \mathrm{Pf}}
\newcommand{\pf}{\text{Pf} \,}
\newcommand{\pfr}{\text{Pf}^\prime \,}

\newcommand{\be}{\begin{equation}\label}
\newcommand{\ee}{\end{equation}}
\newcommand{\bea}{\begin{eqnarray}\label}
\newcommand{\eea}{\end{eqnarray}}

\newcommand{\la}{\langle}
\newcommand{\ra}{\rangle}

\newtheorem{thm}{Theorem}

\newtheorem{propn}{Proposition}[section]

\begin{document}

\title{New Ambitwistor String Theories}
\author{Eduardo Casali$^\ddagger$, Yvonne Geyer$^\dagger$, Lionel Mason$^\dagger$, Ricardo Monteiro$^\dagger$ \\  \& Kai A.~Roehrig$^\ddagger$. \vspace{.3cm}
\\ \small{$^\dagger$Mathematical Institute, University of Oxford, Woodstock Road, Oxford OX2 6GG, UK
\\ \vspace{.1cm}
$^\ddagger$DAMTP, University of Cambridge, Wilberforce Road, Cambridge CB3 0WA, UK}}

\abstract{We describe new ambitwistor string theories that give rise  to the recent amplitude formulae for Einstein-Yang-Mills, (Dirac)-Born-Infeld, Galileons and others introduced by Cachazo, He and Yuan.  In the case of the Einstein-Yang-Mills amplitudes, an important role is played by a novel worldsheet conformal field theory that provides the appropriate colour factors precisely without the spurious multitrace terms of earlier models that had to be ignored by hand.  This is needed to obtain the correct multitrace terms that arise when Yang-Mills is coupled to gravity.}


\maketitle

\section{Introduction}
Witten's twistor-string theory \cite{Witten:2003nn} led to remarkably simple formulae for tree-level gauge theory amplitudes in four dimensions \cite{Roiban:2004yf}.  Subsequently, the development of new tree-level formulae have led the way towards finding the underlying worldsheet models.  An analogous formula for $N=8$ supergravity  was discovered \cite{Cachazo:2012kg,Cachazo:2012pz} and led to the construction of a twistor-string theory for $N=8$ supergravity \cite{Skinner:2013xp}.  Formulae in arbitrary dimensions were then discovered by Cachazo, He and Yuan (CHY) for Einstein and Yang-Mills amplitudes \cite{Cachazo:2013iea,Cachazo:2013hca,Cachazo:2013gna} based on the scattering equations (which underpin all these formulae including the original twistor string \cite{Witten:2004cp}).  These in turn led to the discovery of a family of `ambitwistor-string' theories \cite{Mason:2013sva,Berkovits:2013xba}. 
These are chiral infinite tension string theories that provide a natural generalization of twistor-string theories from four dimensions to arbitrary dimension.  They give the physical theory underlying  these formulae and lead to new extensions.  For example, in the case of the critical type II gravity model in 10 dimensions, the ambitwistor model leads to proposals for how the formulae might be extended to loop amplitudes \cite{Adamo:2013tsa,Casali:2014hfa,Adamo:2015hoa}. They also give new insights into the relationship \cite{Strominger:2013lka, Strominger:2013jfa} between asymptotic symmetries and soft theorems \cite{Adamo:2014yya,Geyer:2014lca,Lipstein:2015rxa,Adamo:2015fwa}.

\begin{figure}
\begin{center}

\tikzstyle{block} = [rectangle, draw, fill=blue!20,
    text width=3em, text centered, rounded corners, minimum height=3em]
\tikzstyle{line} = [draw, thick, -latex']
\tikzstyle{cloud} = [draw, ellipse,fill=red!20, node distance=1.6cm,
    minimum height=3em]

\begin{tikzpicture}[node distance = 2cm, auto]
    \node [cloud] (gr) {Gravity: 
    };
    \node [block, below of=gr] (em) {EM
    };
     \node [block, below of=em] (eym) {EYM
     };
      \node [block, right of=eym, node distance=4.5cm] (ym) {YM
      };
       \node [block, below of=ym] (yms) {YMS
       };
     \node [block, below of=yms] (gyms) {gen.\ YMS 
     };
     \node [block, right of=gr, node distance=4.5cm] (bi) {BI};
     \node [block, below of=bi] (dbi) {DBI};
     \node[block, left of=yms, node distance=4.5cm] (phi4) {$\phi^4$};
     \node[block, right of=ym, node distance=4.6cm] (new) {NLSM};
    \path [line] (gr) --node {compactify} (em);
    \path [line, dashed] (em) --node {generalize}   (eym);
    \path [line] (ym) --node {compactify}   (yms);
     \path [line, dashed] (yms) --node {generalize}   (gyms);
    \path [line, densely dotted] (gr) --node {``compactify''}   (bi);
        \path [line, densely dotted] (em) --node {``compactify''}   (dbi);
      \path [line] (bi) --node {compactify}   (dbi);
       \path [line] (eym) --node {single trace} (ym);
        \path [line] (yms) --node {corollary}   (phi4);
        \path[line, densely dotted] (ym) --node [above] {``compactify''} (new);
 \path [line, dashed] (gr)--++ (-1.2cm, 0cm) |- node [near start, left]{squeeze} (eym);
  \path [line, dashed] (ym)--++ (2.2cm, 0cm) |- node [near start, right]{squeeze} (gyms);
\end{tikzpicture}
\caption{Theories studied by CHY and operations relating them.}
\label{fig:intro}
\end{center}
\end{figure}
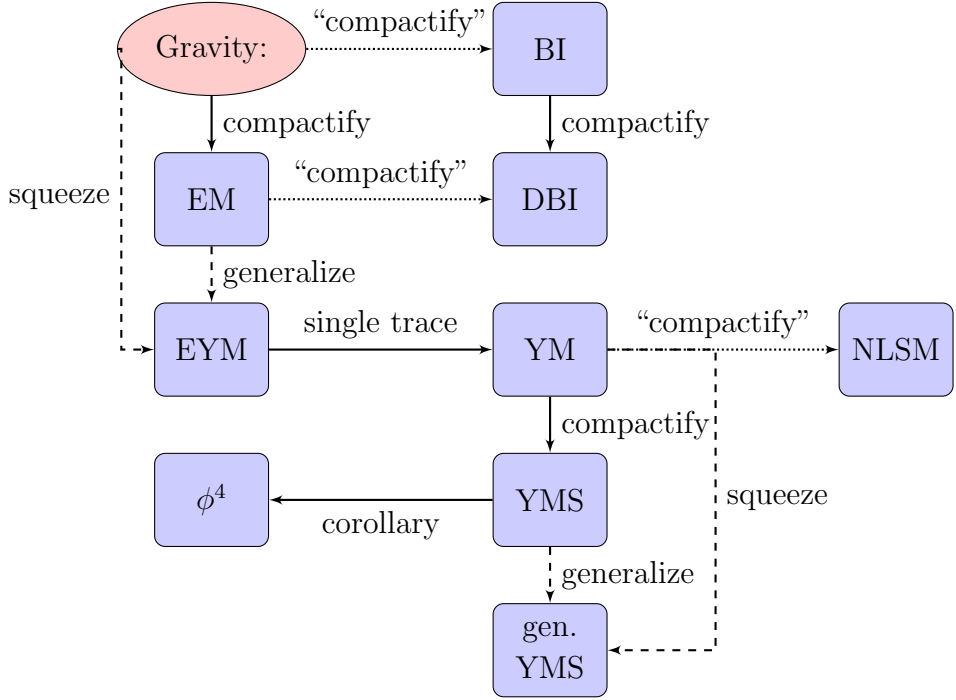

Recently, CHY discovered new formulae for a large collection of theories.  These include the original Einstein (E), Yang-Mills (YM) and biadjoint scalar (BS), together with new formulae for Einstein Maxwell (EM), Einstein Yang-Mills (EYM), (Dirac)-Born-Infeld ((D)BI), Galileons (G), Yang-Mills Scalar (YMS) and nonlinear sigma model (NLSM) \cite{Cachazo:2014nsa,Cachazo:2014xea}, see  figure \ref{fig:intro} below for their diagram\footnote{We thank CHY for permission to reproduce their diagram.} of new amplitude formulae and relationships between them.  They raised the challenge to find the underlying ambitwistor string theories that give rise to these formulae. Indeed, Ohmori in parallel work has already found the ambitwistor strings for the BI and Galileon theories  \cite{Ohmori:2015sha}.  It is the purpose of this paper to review and to give further details for his results  and to show how the remaining new CHY formulae also arise from new ambitwistor string models.

The formulae of Cachazo, He and Yuan give rise to formulae for scattering amplitudes as a sum over solutions to the scattering equations 
of certain  Pfaffian and determinant expressions.  They give tree amplitudes in the general form of an integral over $n$ copies of the Riemann sphere
$$
\cM(1,\ldots,n)=\delta^d\left(\sum_i k_i\right)\int_{(\CP^1)^n} I^l I^r \frac{ \prod_i \bar \delta(k_i \cdot P(\sigma_i)) d\sigma_i}{\rm{Vol} \, G}
$$
where $\sigma_i$, $i=1,\ldots,n$ are complex coordinates on each of the Riemann spheres, $k_i$ the null momenta of the massless particles in the scattering process, and
$$
P^\mu(\sigma)=\sum_{i=1}^n \frac{k^\mu_i}{\sigma-\sigma_i}\, ,
$$
and $G=SL(2,\C)\times \C^3$ is the residual gauge symmetry of the ambitwistor string fixed according to the standard Fadeev-Popov procedure.  The integrand naturally decomposes into factors   
 $I^l$ and $I^r$ that depend on the $\sigma_i, k_i$ and the polarization and/or colour data of the particles whose scattering is being computed and depends on the theory.  The delta functions 
 $$
\bar{\delta}(z)=\dbar\frac{1}{2\pi i z}=\delta(\Re z)\delta (\Im z) d\bar{z},
$$
where $\Re$ denotes the real part and $\Im$ the imaginary part, impose the {\em scattering equations}: $k_i\cdot P(\sigma_i)=0$.  The integrals essentially reduce to a sum over solutions to the scattering equations of $I^lI^r$ multiplied by a Jacobian factor.  The $I^l$ and $I^r$ can be chosen from  five different choices  and the various theories arise from the different possible combinations.  

Ambitwistor strings are chiral infinite tension analogues of RNS strings that can be interpreted, after reduction of constraints, as strings whose target space is the space of complexified null geodesics in Minkowski space.  This space of complexified null geodesics has become known as ambitwistor space. Ambitwistor strings are built out of a basic bosonic model together with worldsheet matter.  The bosonic model leads to a framework in which the  vertex operators required for amplitude calculations incorporate the scattering equations.  The vertex operators also allow for the insertion of two currents $v^l$ and $v^r$ and these can be constructed from additional worldsheet matter (the natural choice for $v^l$ and $v^r$ in the bosonic model does not seem to lead to interesting amplitudes).     
The various Pfaffians, determinants or Parke-Taylor factors that are possible choices for the $I^l$ and $I^r$ arise as worldsheet correlators of currents for  the $v^l$ and $v^r$ respectively.   Corresponding to the five choices for the $I^l$ and $I^r$ in the CHY formulae we will introduce five choices of worldsheet matter, see table \ref{models0}.

\renewcommand{\arraystretch}{2}
\begin{table}[t]{\small
\begin{tabular}{|c||l|l|l|l|l|}
  \hline
  \diagbox{$S^l$}{$S^r$}& $S_\Psi$ & $S_{\Psi_1,\Psi_2}$ & $S_{\rho,\Psi}^{(\widetilde m)}$ & $S_{YM,\Psi}^{(\widetilde N)}$ & $S_{YM}^{(\widetilde N)}$\\ \hline \hline
  $S_\Psi$ & E &  & &  & \\ \hline 
  $S_{\Psi_1,\Psi_2}$ & BI & Galileon &  &  & \\ \hline  
  $S_{\rho,\Psi}^{(m)}$ & $\stackrel[\text{U}(1)^{m}]{}{\text{EM}}$ & DBI & $\stackrel[\text{U}(1)^{m}\times \text{U}(1)^{\widetilde m}]{}{\text{EMS}}$
   &  & \\ \hline 
  $S_{YM,\Psi}^{(N)}$ & EYM & ext.\ DBI & $\stackrel[\text{SU}(N)\times \text{U}(1)^{\widetilde m}]{}{\text{EYMS}}$ & $\stackrel[\text{SU}(N)\times \text{SU}(\widetilde N)]{}{\text{EYMS}}$ & \\ \hline  
  $S_{YM}^{(N)}$ & YM &  Nonlinear $\sigma$ & $\stackrel[\text{SU}(N)\times \text{U}(1)^{\widetilde m}]{}{\text{EYMS}}$ & $\stackrel[\text{SU}(N)\times \text{SU}(\widetilde N)]{}{\text{gen YMS}}$ & $\stackrel[\text{SU}(N)\times \text{SU}(\widetilde N)]{}{\text{Biadjoint Scalar}}$\\ \hline 
 \end{tabular}}
 \caption{Theories arising from the different choices of matter models.} \label{models0}
\end{table}

In the original models of \cite{Mason:2013sva} just two ingredients were used to construct $I^l$ and $I^r$,  worldsheet supersymmetry $S_\Psi$, and a current algebra $S_J$.   Einstein, Yang-Mills and Biadjoint scalar theories were obtained from  the choices $(S^l,S^r)= (S_\Psi,S_\Psi), (S_\Psi,S_J)$ and $(S_J,S_J)$ respectively.  The current algebra $S_J$ has the defect that it also leads to multi-trace terms in its correlators that were ignored by hand.  Here we use a different worldsheet CFT, the comb system\footnote{This was originally introduced by David Skinner and one of us \cite{Casali:2013} in the context of twistor-strings, but never published.}, $S_{CS}$.  This gives a new way to obtain colour factors together with their Parke-Taylor cyclic denominators in such a way that these multi-trace terms simply do not appear. Furthermore, the colour factors are presented not as cyclic single trace terms, but as strings of structure constants arranged in a `comb', hence the name.  However, the number of gauge particles in this system is doubled. To remedy this issue, a reduced system $S_{YM}$ with the correct number of gauge particles can be constructed, but this is is always anomalous. Nevertheless, it is sufficient to produce the correct tree amplitudes and so we use this system instead of the current algebra in the table \ref{models0}.  It can be replaced by $S_{CS}$ if we are seeking an anomaly-free theory, but then we must accept the doubling of gauge particles.  The remaining systems that we use will be combinations of these (with $S_{\rho,\Psi}$ essentially being the abelian limit of the combination $S_{CS,\Psi}$ of the comb system with worldsheet supersymmetry).  These will be described in more detail in the main text.

There are a number of questions that one can ask about these models.  For example, if they are critical and anomaly free, then one can attempt to calculate loop amplitudes by taking the correlation functions on higher genus Riemann surfaces as described in \cite{Adamo:2013tsa}.  For this to work at 1-loop, we must check modularity.  Another issue is as to whether there are any further vertex operators in the theories and if so, we can hope to extend the theory to include additional fields and calculate the corresponding amplitudes.  This in particular happens in theories containing worldsheet supersymmetry $S_\Psi$ and leads to supersymmetric extensions of the theories and amplitudes as described also in \cite{Adamo:2013tsa}.  We will find a number of new critical theories and give a brief discussion of these issues in the conclusions section.

Potentially the most interesting of these models is that for Einstein-Yang-Mills.  We obtain these in two forms.  One gives the correct CHY tree-level amplitudes,  but is anomalous using $S_{YM}$.  The other has vanishing central charge in 10 dimensions, but has doubled gluons in the theory.  The gauge theory part of the action is given by
\begin{equation}
\label{eq:tYM0}
S_{T^*\text{YM}} = \int d^D x \;\tr( a_\mu\, D_\nu F^{\mu\nu}) \, , 
\end{equation}
and we refer to it as $T^*$YM as it describes a linearized Yang-Mills field $a$ propagating on a full Yang-Mills background for the field $A$ with curvature $F$.  Here $a$ is canonically conjugate to $F$ hence the name $T^*$YM as opposed to $T$YM.  This should give correct Yang-Mills amplitudes at one loop but has no higher loop amplitudes in the pure gauge sector. In its critical dimension $d=10$, we would expect it to give a valid expression for the 1-loop integrand for Yang-Mills also.

Table \ref{models0}, showing how the theories are determined in terms of a pair of worldsheet systems, is a remarkable manifestation of the notion of double copy. This notion has been explored mostly in the context of gravity amplitudes, which are obtained as the double copy of gauge theory ones \cite{Kawai:1985xq,Bern:2008qj}. In the formalism of the scattering equations, this is the double copy of Pfaffian factors, and in ambitwistor string theory, this is the double copy of the worldsheet system $S_\Psi$, as in table \ref{models0}. The amplitude formulae of ref.~\cite{Cachazo:2014xea} and our results extend this notion to a range of other theories. Regarding the relation to previous work, we should mention that a double copy construction for Einstein-Yang-Mills amplitudes was first presented in \cite{Bern:1999bx} for the single trace contribution, and in \cite{Chiodaroli:2014xia} for the complete amplitude, with results also at loop level. These double copy constructions are based on the colour-kinematics duality \cite{Bern:2008qj,Bern:2010ue}, whose relation to the scattering equations has been explored in \cite{Cachazo:2013iea,Monteiro:2013rya,Naculich:2014rta}.

\section{Ambitwistor string models}

\subsection{The bosonic ambitwistor string}
The original bosonic ambitwistor string starts with an action obtained by complexifying the standard massless superparticle.
We complexify the worldline to be a Riemann surface $\Sigma$ with holomorphic coordinate $\sigma\in\C$, and  the target space is taken to be the cotangent bundle $T^*\C^d$ of complexified space-time (so that the $(P_\mu,X^\mu)$, $\mu,\nu=1,\ldots d$ are holomorphic coordinates and space-time has a holomorphic flat metric $\eta_{\mu\nu}$).  We use  the bosonic action
\be{boson-str}
S_B=S_B[X,P]=\frac1{2\pi}\int_\Sigma  P\cdot \dbar X +  \tilde e P\cdot  P\, ,
\ee
where 
$\dbar X = \rd\bar\sigma\, \partial_{\bar\sigma} X$ and the `$\cdot$' will be used to denote contraction of indices whenever the contraction is unambiguous. We take the components of  $P_\mu$ to be complex (1,0)-forms on the worldsheet, so that  suppressing space-time indices, $P = P_{\sigma}(\sigma)\rd\sigma\in K:=T^*\Sigma$. It follows that $\tilde e$ must be a (0,1)-form on $\Sigma$ with values in $T:=T\Sigma$ the holomorphic tangent bundle.  It plays the role both of a Lagrange multiplier that enforces $P^2=0$ and of a gauge field for the transformation
$$
\delta \tilde e= \dbar \alpha\, , \qquad \delta X= \alpha P\, , \qquad \delta P=0\, .
$$
Restricting to $P^2=0$ and reducing by the gauge freedom reduces us to ambitwistor space, the space of complex null geodesics.   

This gauge freedom can be fixed by setting $\tilde e=0$ with the introduction of ghosts $(\tilde b,\tilde{c})\in (K^2,T)$, where $K=T^*\Sigma$ is the bundle of $(1,0)$-forms.
The usual diffeomorphism freedom can also be parametrized using $\dbar_e=\dbar+e \partial$ and fixed with $e=0$, together with the introduction of the usual ghosts $(b,{c})\in (K^2,T)$.  This leads to the basic BRST operator
$$
Q_B=\oint cT+\tilde cP^2\, .
$$

It was shown in \cite{Mason:2013sva} (see also \cite{Adamo:2013tsa} for a more systematic treatment) that these structures are sufficient to lead to a framework in which amplitudes are computed using vertex operators built from some operator $V\in K^2$.  This will be the case for all our ambitwistor-string theories. When $\Sigma=\CP^1$, there are three zero-modes each for $c$ and $\tilde c$, and three vertex operators are fixed,  coming as $c_i\tilde c_i V_i \e^{ik_i\cdot X}$ at $\sigma_i$, $i=1,2,3$.  The remaining $n-3$ are integrated
$$
\mathcal{V}_i=\int \bar\delta(k_i\cdot P(\sigma_i)) \e^{ik_i\cdot X} V_i(\sigma_i)\, ,
$$
where
$$
P(\sigma)=\sum_j\frac{k_j}{\sigma-\sigma_j}, \qquad \quad \bar\delta(z)=\dbar\frac{1}{2\pi i z}=\delta(\Re z)\delta(\Im z) d \bar z \, .
$$
The delta-functions $\bar\delta(k_i\cdot P(\sigma_i))$ impose the scattering equations that provide the backbone of all the  CHY formulae and their twistor-string precursors.  We will take these aspects of the scattering amplitude calculations for granted in the following and will not mention them further.

\subsection{Vertex operators and worldsheet matter}

In general, we will consider theories with actions of the form 
$$
S_B+S^l+S^r
$$
where  $S^l$ and $S^r$ are distinct matter theories on $\Sigma$ that will be described in more detail in the next two sections.  These will contribute to
 our formulae for the vertex operators $V\in K^2$.  The vertex operators all have an $\e^{ik\cdot X}$ factor with the remainder $V$ factorizing into two independent currents,
$$
V=v^lv^r \, , \qquad v^l, v^r \in K.
$$
The $v^l$, $v^r$ will be  constructed from the matter models,  $S^l$ and $S^r$ respectively and constrained by quantum consistency, BRST invariance, and perhaps further discrete symmetries.    Invariance under $Q_B$ for example implies $k^2=0$ because the $P^2$ term in the BRST operator brings down $k^2$ in its double contraction with $\e^{ik\cdot X}$.  We will also use the notation $u^l$, $u^r$  for such currents when they are fixed with respect to fermionic symmetries, see below.  

Essentially the only candidate for $v^l$ and $v^r$ in the purely bosonic model above is $\epsilon\cdot P$ for some polarization vector $\epsilon^\mu$ defined up to multiples of $k^\mu$ under $Q_B$ equivalence. This leads to unphysical formulae for gravity amplitudes, or at least with no clear interpretation.  In order to obtain more interesting models,  we will introduce worldsheet matter models $S^l$ and $S^r$ that will generate the currents $v^l$ and $v^r$ in the vertex operators.  In general, we will take the models $S^l$ and $S^r$ to be distinct matter theories so that the correlator will factorize into a product of one for the left currents and one for the right currents, and we will be able to calculate them separately.  In order to ensure that the only allowed vertex operators do indeed factorize in this way, we will impose discrete symmetries that are analogues of the GSO symmetries of conventional string theories, and we will use that name for these symmetries as well.

 \section{Worldsheet matter models and their correlators}
 In \cite{Mason:2013sva}, two matter models were considered: (1) $S_\rho$, a current algebra which we will take to be generated by free fermions, and (2) $S_\Psi$, which introduces a degenerate worldsheet supersymmetry.  This latter extends $Q$ so as to change the choice of current $v=\epsilon\cdot P$ in the bosonic model to one that we will want.  These led to three models with $(S^l,S^r)$ given by $(S_{\Psi_l},S_{\Psi_r})$ for type II supergravity, $(S_\Psi,S_\rho)$ for Yang-Mills amplitudes and $(S_{\rho_l},S_{\rho_r})$ for amplitudes of a biadjoint scalar theory.  In this paper we will consider a third type of matter that we call the `comb system'  $S_{CS}$, \cite{Casali:2013}, a worldsheet conformal field theory that will be important for Yang-Mills amplitudes so called because its correlators give colour invariants in the form of comb structures built out of structure constants rather than colour traces.  In the rest of this section, we describe these matter systems, and the natural currents to which they give rise as candidates for $v^l$ and $v^r$ and their correlation functions.  In the next section we see how these are altered when these systems are combined.

\subsection{Free fermions \texorpdfstring{$S_\rho$}{S-rho} and current algebras \texorpdfstring{$S_j$}{S-j}.}  
 The standard action for `real' free fermions $\rho^a\in K^{1/2}$, $a=1,\ldots m$, is
$$
S_\rho=\int \rho^a\dbar \rho^a\, ,
$$
(the summation convention is assumed).  The term `real' is used to distinguish them from the complex fermion system given by 
$$
S_{\rho,\tilde\rho}=\frac1{2\pi i} \int \tilde\rho_a \dbar\rho^a.$$  The simplest currents in the real case are $j^{ab}=\rho^a\rho^b$ and form an elementary example of a current algebra for $SO(m)$ (in the complex case $j^a_b=\tilde{\rho}_b\rho^a$ generate a current algebra for $\SU(m)$).

More generally, we can consider an arbitrary current algebra $j^a\in K\otimes \g$, where $\g$ is some Lie algebra, $a=1,\ldots,\dim \g$, satisfying the usual current algebra OPE,
\be{current-alg}
j^a(\sigma)j^b(0)\sim \frac{k\delta^{ab}}{\sigma^2} + \frac{if^{abc}t^c(\sigma)}{\sigma}+\ldots\, ,
\ee
where $f^{abc}$ are the structure coefficients, $[t^a,t^b]=f^{abc} t^c$, $\delta^{ab}$ is the Killing form,  and $k$ is the level. This could be contructed from free fermions, WZW models or some other construction and we will generally represent such matter as $S_j$.

Given choices of $t\in \g$, the current algebra can contribute $$v=t\cdot j$$ to one or both  factors $v_l$ and $v_r$ of the vertex operators $V$. The current correlators $\la t_1\cdot  j_1 \ldots t_n\cdot  j_n\ra$,  where $j_i=j(\sigma_i)$, lead to Park-Taylor factors:
$$
PT(1,\ldots,n)=\frac{\tr(t_1\ldots t_n)}{\sigma_{12}\sigma_{23}\ldots \sigma_{n1}} \,,
$$
where $\sigma_{ij}=\sigma_i-\sigma_j$.  However, the correlators also lead to multi-trace terms that are ultimately problematic and unwanted.

\subsection{Worldsheet suspersymmetry \texorpdfstring{$S_\Psi$}{S-Psi}.} 

Worldsheet supersymmetry is introduced by adding fermionic worldsheet spinor fields $\Psi^\mu\in \C^d\otimes K^{1/2}$, and a gauge field $\chi \in \Omega^{(0,1)}(T^{1/2})$ for the supersymmetry.   Their action is
$$
S_\Psi= \frac{1}{2\pi i}\int \Psi\cdot \dbar \Psi + \chi P\cdot \Psi\, .
$$
The constraint leads to worldsheet gauge transformations
$$
\delta \chi=\dbar \eta \, , \qquad \delta X=\eta \Psi\, , \qquad \delta \Psi=\eta P\, , \qquad \delta P=0\, ,
$$
where $\eta $ is  a fermionic parameter. Gauge fixing leads to bosonic ghosts $\gamma \in T^{1/2}$ and corresponding antighosts $\beta$.  The BRST operator acquires an extra term
$$
Q_\Psi=\oint \gamma \cG_{\Psi}\, , \qquad \cG_{\Psi}:= P\cdot \Psi\, .
$$ 
On $\CP^1$, the ghosts $\gamma$ have two zero modes.  Thus, as far as the fermionic symmetry is concerned, we need two fixed vertex operators with one current factor  of the form $\delta(\gamma) $ multiplied by a field now with values in $ K^{1/2}$, and then the `integrated' ones (in the fermionic sense) arising from descent.   The relevant currents are
$$
u=\delta(\gamma)\epsilon\cdot \Psi\, , \qquad v=\epsilon\cdot P + k\cdot\Psi\epsilon\cdot\Psi\, ,
$$
with just two of the $u$s required in a correlator.  

These operators are invariant under the discrete symmetry that changes the sign of $\Psi$, $\chi$ and the ghosts.  Imposing invariance under this symmetry will exclude mixing between the ingredients of these operators thought of as parts of $S^l$ and others that might be part of $S^r$.  We will refer to this as GSO symmetry.

The correlators of these currents lead to the reduced Pfaffians of CHY:
$$
\la u_1  u_2 v_3\ldots v_n\ra = \mathrm{Pf}'(M)=\frac{1}{\sigma_1-\sigma_2}\mathrm{Pf}(M_{12})\, ,
$$
where $M$ is the skew $2n\times 2n$ matrix with $n\times n$ block decomposition
$$
M=\begin{pmatrix}
A&-C^T\\C&B
\end{pmatrix} \, , \qquad A_{ij}=\frac{k_i\cdot k_j}{\sigma_{ij}}\, , \qquad B_{ij}=\frac{\epsilon_i\cdot \epsilon_j}{\sigma_{ij}}\, , 
$$
and
$$
C_{ij}=\frac{\epsilon\cdot k_j}{\sigma_{ij}}\, , \;i\neq j, \qquad C_{ii}= -\epsilon_i\cdot P(\sigma_i)\, ,
$$
and $M_{12}$ is $M$ with the first two rows and columns removed.

\subsection{Comb system \texorpdfstring{$S_{CS}$}{S-CS}.}\label{sec-CS}
The comb system  \cite{Casali:2013} was introduced as a way of obtaining colour factors as sequences of contractions of structure constants rather than as colour ordered traces. In general, such contractions can be generated from trivalent diagrams with the structure constants $f^{abc}$ of some Lie algebra at the vertices and contractions $\delta^{ab}$ along the internal edges.  It is well known that these are linearly dependent as a consequence of the Kleiss-Kuijf relations with a basis being given by `combs', with $n-2$ vertices lined up in a row \cite{KK1989,DelDuca:1999rs} and end points given by 1 and $n$:
\vspace{.4cm}
\begin{center}
\includegraphics[width=5cm]{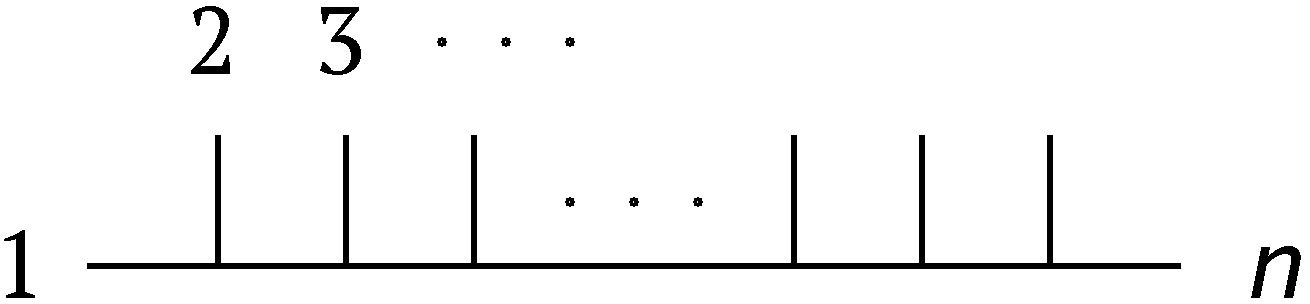} \qquad $\to \qquad f^{a_1a_2b_1}f^{b_1a_3b_2}\cdots f^{b_{n-3}a_{n-1}a_n}$.
\end{center}
\vspace{.4cm}
The comb system has the remarkable property that, in conjunction with worldsheet supersymmetry, only these combs arise from correlators  and not the  multitrace terms that arise from an ordinary current algebra. This system arises from an action for matter fields 
$\rho, \tilde \rho, q, y \in \mathfrak{g}\otimes K^{1/2}$ i.e., worldsheet spinors taking values in the Lie algebra $\mathfrak{g}$ of some gauge group.   The worldsheet action is
$$
S_{CS}=\int  \tilde\rho\cdot \dbar \rho + q\cdot \dbar y + \chi \rho\cdot\left(\frac{\scriptstyle 1}{\scriptstyle 2} [\rho,\tilde\rho]+ [q,y]\right)
$$
 with $\rho, \tilde \rho$ fermionic and $q,y$ bosonic and the $\cdot $ is used to denote the Killing form on the Lie algebra.  As before, $\chi$ is a gauge field on the worldsheet with values in $T^{1/2}\otimes \Omega^{0,1}$ and we are gauging 
the current\footnote{With different assignment of worldsheet spins this current would be a normal BRST current.  If we were to take $\rho, y$, scalars and $\tilde{\rho}, q$  sections of $K$, then $\rho$ and $\tilde\rho$ could be taken to be the ghosts associated to  gauge fixing a worldsheet gauge field $a\in\Omega^{0,1}\otimes \mathfrak{g}$ with action $\int_\Sigma q\cdot \dbar y +q\cdot [a,y]$.  This fact allows one to see the consistency of this current reasonably rapidly. } $\rho\cdot\left(\frac{\scriptstyle 1}{\scriptstyle 2} [\rho,\tilde\rho]+ [q,y]\right)$, which is a section of $K^{3/2}$.  
The gauging introduces transformations now for fermionic  $\alpha\in T^{1/2}$
$$
\delta(\rho,\tilde\rho,q,y)=\alpha (\frac{1}{2}[\rho,\rho], [\rho,\tilde{\rho}], [\rho,q],[\rho,y])\, , \qquad \delta \chi=\dbar\alpha\, . 
$$ 
 As in the case of worldsheet supersymmetry, gauge fixing gives bosonic ghosts $\gamma\in T^{1/2}$ and antighosts $\beta$ with a contribution to the BRST operator of
 $$
Q_{CS}=\oint \gamma  \cG_{CS}\, , \qquad \cG_{CS}:=\rho\cdot\left(\frac{\scriptstyle 1}{\scriptstyle 2} [\rho,\tilde\rho]+ [q,y]\right)\, .
$$

As for $S_\Psi$, there are two zero-modes for the ghosts, and so we will need two fermionically fixed operators with the rest integrated.  The currents that contribute to the vertex operators in this system now depend on a Lie algebra element $t\in\g$,  with two types of fixed and integrated ones respectively being
$$
u=\delta(\gamma)t\cdot \rho\, , \quad \tilde u=\delta(\gamma)t\cdot \tilde \rho\, , \qquad  v=\half t\cdot [\rho,\rho]\, , \quad \tilde v=t\cdot\left( [\rho,\tilde\rho]+ [q,l]\right)\, .
$$
Here $v=\{Q_{CS},u\}$ and $\tilde v=\{Q_{CS},\tilde u\}$, and, in any correlator, we need two fixed and the remaining unfixed vertex operators\footnote{
A  more symmetric way to understand this is to say that we choose all unintegrated vertex operators, but then we must insert $n-2$ `picture-changing operators'
$$
\Upsilon=\delta(\beta) \rho\cdot\left(\frac{\scriptstyle 1}{\scriptstyle 2} [\rho,\tilde\rho]+ [q,y]\right)\, .
$$
These could be inserted anywhere in general.  If inserted at one of the $u, \tilde u$ insertion points, it will convert it into a corresponding $v,\tilde v$. A similar approach can be taken for correlators  associated with the $S_\Psi$ matter system. 
}.  
Notice that $\tilde v= t\cdot j$, where $j^a$ is a level zero current algebra, and that
$$
\tilde v (\sigma)\; t'\cdot \tilde \rho (0) \sim - \frac{[t,t']\cdot \tilde \rho (0)}{\sigma} + \ldots \,,
\qquad \tilde v (\sigma)\; t'\cdot \rho (0) \sim - \frac{[t,t'] \cdot \rho (0)}{\sigma} + \ldots \;.
$$
The correlators are as follows
\begin{propn}[Casali-Skinner] Correlators of the currents $u,v,\tilde u,\tilde v$ are only nonvanishing when there is just one untilded current and give 
$$
\la u_1 \tilde v_2\ldots \tilde v_{n-1} \tilde{u}_n\ra=\la \tilde u_1 v_2 \tilde v_3\ldots \tilde v_{n-1} \tilde{u}_n\ra=
\cC(1,\ldots,n)
$$
where
$$
\cC_n=\cC(1,\ldots,n):=\frac{\tr(t_1[t_2,[\ldots,[t_{n-1} ,t_n] \ldots] ])}{\sigma_{12}\sigma_{23}\ldots \sigma_{n1}} + \mathrm{Perm}(2,\ldots,n-1)\, .
$$
\end{propn}

Instead of giving the colour traces, we obtain `combs', i.e., strings of structure constants $\tr(t_1[t_2,[\ldots,[t_{n-1} ,t_n] \ldots] ])$ as described in \cite{DelDuca:1999ha,DelDuca:1999rs}.  

The argument is as follows. The fact that we can have at most two $u,\tilde u$s is the standard counting of $\gamma$ ghost zero modes. Consider the $\rho, \tilde\rho$ contractions: that these are the only nontrivial correlators comes from the need to have as many $\rho$s as $\tilde \rho$, so it is easily seen that we can have only one untilded current which can either be a $u$ or a $v$.  The $v,\tilde{v}$s connect  along a `comb', whereas the $u,\tilde u$s form the ends.  Such contractions connecting all $n$ vertex operators form the right hand side above.  We can also have contractions in which a collection of $\tilde v$s come together in contractions to form a loop.  This is where the $(q,y)$ system comes into play. These can only form loops, but, being bosonic, their loop contractions cancel such loop contractions from the $\rho, \tilde{\rho}$ system.  This can also be seen from the form of the current algebra generated by the $\tilde v$s.  This has by construction level zero so that, after a sequence of OPE's, cannot generate a nontrivial trace.

\subsection{Other systems with comb structure, \texorpdfstring{$S_{YM}$}{S-YM}.}\label{sec-YM}  A problem with the CS system above is that there are clearly two types of gluon, tilded und untilded corresponding to the vertex operators $(\tilde u,\tilde v)$ and $(u,v)$ respectively.
We will see that this is not appropriate for pure Yang-Mills although it does give a theory that is sufficient to generate Einstein-YM tree amplitudes correctly on certain trace sectors, the ones selected by the choice of untilded operators.\footnote{One may try to symmetrise the correlator in tilded versus untilded gluonic operators, for instance by using $u_t+\tilde u_t$ and $v_t+\tilde v_t$, but then there will be an over-counting of contributions, so that the relative factors of different terms are not correct.} The system we introduce here will give the complete Einstein Yang-Mills amplitude from a single correlator, but will be anomalous.

A worldsheet CFT that will generate YM following the ideas above requires the following ingredients.  We need a fermionic worldsheet spinor $\rho^a\in \g$ for the fixed vertex operator, a current algebra $v^a\in \g$ at level zero for the integrated one; the level zero allows us to avoid multitrace terms and loops.  Finally we need a spin 3/2 current $\cG_{YM}$ with the following OPE to give the appropriate group compatibilities and descent:
\be{requirements}
\rho^a(\sigma)\rho^b(0) \sim \frac{\delta^{ab}}{\sigma}\, , \;\;  v^a(\sigma)\rho^b(0)\sim \frac{f^{abc}\rho^c(0)}{\sigma}\, , \;\; \cG(\sigma) \rho^a(0)\sim \frac{v^a(0)}{\sigma}\, , \;\; \cG(\sigma)\cG(0)\sim 0\, .
\ee
It is easy to see that this can be partially realized with $\rho^a $ a `real' free fermion with action $\frac{1}{2}\int \rho^a\dbar \rho^a$ and with 
$$
v^a= -\frac{1}{2} f^{abc}\rho^b\rho^c + j^a\, ,\quad j^a(\sigma)\rho^b(0)\sim 0\, ,
$$
we will obtain the first two of the equations above.   
In order for $v^a$ to be a current algebra with level zero, because $\frac{1}{2}f^{abc}\rho^b\rho^c$ is  a current algebra with level $-C$ where $f^{abc}f^{\tilde abc}=C\delta^{a\tilde a}$, we must take $j^a$ to be a current algebra with level $k=C$. There are many ways to do this, so let us leave this to one side for a moment.  We then need to construct $\cG$.  In order for $\cG$ to generate $v^a$ from $\rho^a$, we must have
$$
\cG= -\frac{1}{6} f^{abc}\rho^a\rho^b\rho^c + \rho^a j^a +\ldots
$$       
where the $\ldots$ has nonsingular OPE with $\rho^a$ and $j^a$. At this point, however, we see that an anomaly arises preventing $\{\cG,\cG\}=0$. To be specific,
$$
\cG(\sigma)\cG(0) \sim \frac{C\,\textrm{dim}(G)}{\sigma^3} + \frac{:j^aj^a(0):}{\sigma},
$$ 
where we recall that the energy-momentum tensor of the current algebra $j$ is given by $T(\sigma)=:j^aj^a(\sigma):/2k$. Therefore, we are able to satisfy the first three equations of \eqref{requirements}, while the last equation is anomalous.

\subsection{Central charges}

We remark that the theories  $S_B$, $S_\rho$, $S_\Psi$ and $S_{CS}$  above respectively have central charges 
$$
 c_B=2d-52,\qquad  c_\rho =m/2\, , \qquad  c_\Psi= d/2+11, \qquad c_{CS}=11\, ,
$$
the latter being just that of the $\beta-\gamma$ system as the $\dim G$ parts cancel via supersymmetry.  (This can be different if the $(q,y)$ are not taken to be spin $1/2$.)  Notably, the type II supergravity model is critical in 10 dimensions as then $c_B+2c_\Psi=0$.  These considerations are less interesting for $S_{YM}$ as that theory is already quite anomalous, and in any case its central charge will depend on the choice of  current algebra $j^a$.


\section{Combined Matter models}

On their own, the new worldsheet matter theories $S_{CS}$ and $S_{YM}$ of the previous section do little more than give an alternative to the current algebras in the original models of \cite{Mason:2013sva} that avoids the multitrace terms that were neglected by hand.  To obtain new theories, we will consider the contributions to $S^l$ or $S^r$ of combinations of the above matter systems.
Even without $S_{CS}$ and $S_{YM}$, we will obtain a number of interesting new models. 
Here we will consider the allowable vertex operators and the correlators of the various combinations that we can form.  These are summarized in the table \ref{matter-combinations}. 

\begin{table}[h]  {\footnotesize
\begin{tabular}{|l||l|l|l|l|}\hline
  & Fermionic current $\mathcal{G}$ & Matter & Vertex operators & Correlator\\ \hline\hline
 $S_\Psi$ & $P\cdot \Psi$ & $\Psi$ & $u_\Psi=\delta(\gamma)\,\epsilon\cdot\Psi$ & Pf$'(M)$\\ \hline
 \multirow{2}{*}{$S_{\Psi_1,\Psi_2}$} & $P\cdot \Psi_1$ & \multirow{2}{*}{$\Psi_1,\,\Psi_2$} & $u_{\Psi_1}=\delta(\gamma_2)\,k\cdot\Psi_1$ & \multirow{2}{*}{$\big($Pf$'(A)\big)^2$}\\
 & $P\cdot \Psi_2$ & & $u_{\Psi_2}=\delta(\gamma_1)\,k\cdot\Psi_2$ & \\ \hline
 \multirow{2}{*}{$S_{\rho,\Psi}$} & \multirow{2}{*}{$P\cdot \Psi$} & \multirow{2}{*}{$\Psi$, $\rho_a \quad a=1,\dots,m$} & $u_\Psi=\delta(\gamma)\,\epsilon\cdot\Psi$ & \multirow{2}{*}{Pf$(\chi)\,$Pf$'(M|_{\text{red}})$}\\
 &&& $u_{\rho_a}=\delta(\gamma)\,\rho_a$ & \\ \hline
 \multirow{3}{*}{$S_{CS,\Psi}$} & \multirow{3}{*}{$P\cdot \Psi+$tr$\left(\rho(\frac{1}{2}[\tilde{\rho},\rho]+[q,y])\right)$} & \multirow{3}{*}{$\Psi,\,(\tilde{\rho},\,\rho),\,(q,y)$} & $u_\Psi=\delta(\gamma)\,\epsilon\cdot\Psi$ & \multirow{3}{*}{$\mathcal{C}_{(1)}\dots\mathcal{C}_{(m)}$Pf$'(\Pi)$}\\
 &  & & $\tilde{u}_{CS}=\delta(\gamma)\,$tr$(t\tilde{\rho})$ & \\
 &  & & $u_{CS}=\delta(\gamma)\,$tr$(t\rho)$ & \\ \hline
 \multirow{2}{*}{$S_{CS}$} & \multirow{2}{*}{tr$\left(\rho(\frac{1}{2}[\tilde{\rho},\rho]+[q,y])\right)$} &
 \multirow{2}{*}{$(\tilde{\rho},\,\rho),\,(q,y)$} & $\tilde{u}_{CS}=\delta(\gamma)\,$tr$(t\tilde{\rho})$ & \multirow{2}{*}{$\mathcal{C}_n$}\\
 &  & & $u_{CS}=\delta(\gamma)\,$tr$(t\rho)$ &  \\ \hline
\end{tabular}}
\caption{Table of matter models, their combinations and worldsheet correlators}
\label{matter-combinations}
\end{table}

\subsection{\texorpdfstring{$S_{\rho,\Psi}$}{S-rho,Psi}} 
Here we take 
$$
S^l=S_{\rho,\Psi}:=S_\rho+S_\Psi\, .
$$   
Although the free fermion system $S_\rho$ would seem to naturally lead  to  the  $SO(m)$ current algebra $j^{ab}=j^{[ab]}=\rho^a\rho^b$,  and therefore superficially be thought to give the same results as the current algebra, in the presence of worldsheet supersymmetry, the currents $j^{ab}$ as constituents of vertex operators are not BRST invariant, since
$$
\{Q_\Psi, j^{ab} \e^{ik\cdot X}\}= ik\cdot\Psi j^{ab} \e^{ik\cdot X}\neq 0\, .
$$
However, in this context allowable fixed and integrated currents are respectively
$$
u^a=\delta(\gamma) \rho^a\, , \qquad v^a=k\cdot \Psi \rho^a\, , \quad a=1,\ldots , m\, .
$$
We also have the standard BRST invariant currents from $S_\Psi$, which in this context we will denote $u_\epsilon=\delta(\gamma)\epsilon\cdot \Psi$ and $v_\epsilon=\epsilon\cdot P+ k\cdot\Psi \epsilon\cdot\Psi$.

In general we will be concerned with a correlator $\la u_1 u_2 v_3 \ldots v_n\ra $ where, if $(\gamma,h)$ is a partition of $1,\ldots , n$,  for  $i\in \gamma $ the current will be one of the new photon currents, and for $i\in h$ it will be a $S_\Psi$ current depending on a polarization vector $\epsilon_\mu$.   The correlator will factorize into one for the constituent $\rho^a$s and one for the $\Psi$s.  We compute these as Pfaffians of the associated matrices of possible contractions in the correlator.  The simplest is the $\rho$ system. If we restrict it to take values in an algebra with vanishing structure constants, e.g. $\oplus^m  \mathfrak{u}(1)$, the OPEs lead to the $|\gamma| \times |\gamma|$ CHY matrix 
$$
\cX_{ij}=\frac{\delta^{a_ia_j}}{\sigma_{ij}}\, , \quad i,j\in \gamma, \quad i\neq j, \quad  \mbox{ otherwise } \quad \cX_{ij}=0\, .
$$
The Kronecker delta $\tr ( t^{a_i}t^{a_j} )$ in the numerator ensures only photons of the same flavour interact.

Much as before, the $\Psi$ system leads to the matrix of possible $\Psi$ contractions
$$
M_{\mathrm{Red}}=\begin{pmatrix}
A^{\gamma\gamma}&A^{\gamma h}& -C^{h\gamma T}\\
A^{h\gamma}& A^{hh}&-C^{hhT}\\
C^{h\gamma}&C^{hh}& B
\end{pmatrix} \,,
$$
where we have divided the matrix into the bock decomposition under $n=|\gamma|+|h|$ and
$$
A_{ij}=\frac{k_i\cdot k_j}{\sigma_{ij}}\, , i\neq j, \quad A_{ii}=0\, ,\qquad B_{ij}=\frac{\epsilon_i\cdot\epsilon_j}{\sigma_{ij}}\, , \quad i,j\in h, i\neq j. 
$$
and 
$$
C_{ij}=\frac{\epsilon_i\cdot k_j}{\sigma_{ij}} \, , \quad i\in h , i\neq j\, .
$$
Finally, the additional $\epsilon\cdot P$ term in the $S_\Psi$ vertex operator is incorporated by setting $C_{ii}=-\epsilon_i\cdot P(\sigma_i)$ as before.  In this case, we obtain a reduced Pfaffian associated with the two fixed vertex operators as before.  Our final correlator expression is therefore
$$
\la u_1 u_2 v_3 \ldots v_n\ra =\Pf(\cX)\Pf'(M_{Red})\,.
$$

Now for the GSO symmetry we require all fields, $\rho, \Psi$ and the ghosts to change sign simultaneously.

\subsection{\texorpdfstring{$S_{\Psi_1,\Psi_2}$}{S-Psi1,Psi2}}  
Here we take two worldsheet supersymmetries 
$$
S^l=S_{\Psi_1}+S_{\Psi_2}\, .
$$
 There are two contributions to the BRST operator, $Q_{\Psi_1}+Q_{\Psi_2}$.    The normal currents from $S_{\Psi_1}$ and  $S_{\Psi_2}$ are no longer invariant as, for example,
 $$
Q_{\Psi_2}\delta(\gamma_1) \epsilon\cdot\Psi_1 \e^{i k\cdot X}= \delta(\gamma_1)  k\cdot\Psi_2 \epsilon\cdot \Psi_1 \e^{i k\cdot X} \neq 0\, .
$$
However, the nontrivial  BRST invariant currents are descendants simply of $\delta(\gamma_1)\delta(\gamma_2)$ so 
$$
u=\delta(\gamma_1) \delta(\gamma_2)\, , \qquad v= k\cdot \Psi_1 k\cdot \Psi_2\, ,
$$
as given in \cite{Ohmori:2015sha} (and we also have partial descendants $\delta(\gamma_1 ) k\cdot \Psi_2$ and $\delta(\gamma_2)k\cdot \Psi_1$). 
Again, the correlator of $n$ such vertex operators factorizes into a product of the  Pfaffians of the matrix of all possible  $\Psi_1$ contractions and that for all $\Psi_2$ contractions.  These matrices are given simply by the $A$ matrix with off-diagonal entries $k_i\cdot k_j/\sigma_{ij}$ as before.  This matrix has co-rank two and we take a reduced Pfaffian (corresponding to the choice of fixed versus integrated vertex operators).  We therefore now obtain
$$
\la u_1u_2v_3v_4\ldots v_n\ra=\Pf'(A)^2\, .
$$

One might ask whether one can carry on to combine three or more $S_\Psi$ systems, but this is not possible on one side, that is, to produce nontrivial BRST invariant currents.

Again for the GSO symmetry we require all fields, $\Psi_1, \Psi_2$ and the ghosts to change sign simultaneously.

\subsection{\texorpdfstring{$S_{YM,\Psi}$}{S-YM,Psi}}

In Sections \ref{sec-CS} and \ref{sec-YM}, we introduced $S_{CS}$ and $S_{YM}$ whose correlators provide the colour  comb-structure  together with Parke-Taylor factors. For the remainder of this section, we will combine each of these two systems with $S_\Psi$. The goal is to obtain the building block of Einstein-Yang-Mills amplitudes that gives the appropriate interactions  between gluons and gravitons. We start by discussing the combined theory $S_{YM,\Psi}$, which is slightly simpler than $S_{CS,\Psi}$ and possesses the main important features. Despite $S_{YM}$ not being quantum-mechanically consistent  - and this problem  extends to $S_{YM,\Psi}$ - we are able to obtain tree amplitudes.  The  theory $S_{CS,\Psi}$ can be made consistent, but has two types of gluons and the corresponding amplitudes arise from an action that is not Yang-Mills (although it contains its classical solutions).

Since both worldsheet matter theories $S_\Psi$ and $S_{YM}$ involve the gauging of spin 3/2 currents $\cG_\Psi=P\cdot \Psi$ and $\cG_{YM}=\rho\cdot\left(-\frac{\scriptstyle 1}{\scriptstyle 6} [\rho,\rho]+ j\right)$, we have the option of gauging both these currents together or separately.  If we gauge them separately, we find that the resulting system is too restrictive to lead to interesting results.  Thus we gauge the sum 
$$
\cG = P\cdot \Psi +\rho\cdot\left(-\frac{ 1}{ 6} [\rho,\rho]+ j\right),
$$
perform gauge fixing and introduce a single set of ghosts $(\beta,\gamma)$. We find that the currents
$$
u_t=\delta(\gamma)\rho\cdot t\, , \quad \quad u_ \epsilon=\delta(\gamma)\epsilon\cdot \Psi
$$
still give us allowed fixed vertex operators. BRST descent leads to the integrated vertex operators
$$
 v_t=k\cdot\Psi \rho\cdot t + v^0_t\,,
 \quad\quad v_\epsilon=\epsilon\cdot P+\epsilon\cdot\Psi k\cdot\Psi,
$$
where $v_t^0$ denotes the original $S_{YM}$ integrated vertex operator, satisfying the OPE relations \eqref{requirements} except the last.  Although the failure of the last relation means that the BRST quantisation is inconsistent, the correlator of the vertex operators does nevertheless  give the correct amplitudes.

In the previous section, we saw that the system $S_{YM}$ on its own gives the correct colour-dressed Parke-Taylor factors, in terms of a comb structure. The combination with $S_\Psi$ leads to additional insertions of $\rho\cdot t$ and these will start additional combs. In this way we obtain multiple colour combs/traces and get the right interactions with gravity states. On the other hand, the system $S_\Psi$ on its own leads to a reduced Pfaffian. The combination with $S_{YM}$ will lead to a different but closely related Pfaffian that incorporates the multi-comb structure. We now describe the complete correlator.

\begin{thm}\label{ym-correlator}  As in \cite{Cachazo:2014xea}, let the sets $g$ index the gluons with vertex operators $u_t,v_t$, and $h$ the gravitons with vertex operators  $u_\epsilon, v_\epsilon$.  To be non-zero, a correlator must contain two fixed vertex operators $u$'s, with the remaining ones being $v$'s. The correlator  is then a sum over all partitions of the gluons into sets $T_1, T_2,\ldots, T_m$, where $\cup_{i=1}^m T_i=g$ and $|T_i|\geq 2$.  Each partition gives rise to the term
\begin{equation}
 \sum_{\substack{c_1 < d_1 \in T_1 \\ \cdots \\ c_m < d_m \in T_m }}  \mathcal{K}(c_1 ,d_1 | T_1) \cdots  \mathcal{K}(c_n ,d_n | T_n)  ~~ \pf ' \left( \begin{array}{ccc|c}
A_{ab} & A_{ a c_j} &  A_{a d_j }  & (-C^T)_{ab}  \\ 
A_{c_i b} & A_{c_i c_j} &    A_{c_i d_j }  & (-C^T)_{c_i b}  \\
  A_{d_i b} &  A_{d_i c_j} & A_{d_i d_j}  &  (-C^T)_{d_i b} \\  \hline
C_{ab} & C_{a c_j} &   C_{a d_j}  & B_{ab} 
\end{array} \right) ~.
\label{eqn:KPf}
\end{equation}
Here, $a,b$ label gravitons and $c_i,d_i$ label gluons in $T_i$, so that $A_{ab}$ is an $|h|\times|h|$ matrix, $A_{c_ib}$ is an $m\times|h|$ matrix, and $A_{c_ic_j}$ is an $m\times m$ matrix. Moreover, we defined
\begin{equation}\label{eqn:Comb-Trace-relation}
\mathcal{K}(i,j | T) =  \sigma_{ji} \; \mathcal{C}(T) ,
\end{equation}
where $\mathcal{C}(T)$ is $\mathcal{C}_n$ restricted to $g\in T$. The reduced Pfaffian $\pf '$ is defined in eq.~\eqref{eq:redpf}.
\end{thm}

The proof is given in \cref{sec:ecomb-proof}. This correlator reproduces the main building block of the CHY formula for Einstein-Yang-Mills amplitudes in \cite{Cachazo:2014xea}.  Although not quite in the same form, the equivalence can easily be seen from Eqs.~(3.16) and (3.17) of \cite{Cachazo:2014xea} and this form is more natural from its derivation as a correlator.

\subsection{\texorpdfstring{$S_{CS,\Psi}$}{S-CS,Psi}}

While $S_{YM,\Psi}$ gives the correct  amplitude, its BRST quantisation is  inconsistent. We can obtain the same structure from $S_{CS,\Psi}$ by combining the worldsheet theories $S_\Psi$ and $S_{CS}$, which has the advantage of being anomaly free but the disadvantage of containing two types of gluons.

As for $S_{YM,\Psi}$ we gauge  the sum of spin 3/2 currents $\cG_\Psi=P\cdot \Psi$ and $\cG_{CS}=\rho\cdot\left(\frac{\scriptstyle 1}{\scriptstyle 2} [\rho,\tilde\rho]+ [q,y]\right)$, introducing the action 
$$
S_{CS_\Psi}=\int \Psi\cdot\dbar\Psi + \tilde \rho\cdot \dbar \rho + q\cdot \dbar y +\chi\left(P\cdot \Psi+ \rho\cdot\left(\frac{\scriptstyle 1}{\scriptstyle 2} [\rho,\tilde\rho]+ [q,y]\right)\right).
$$
Now the Lie-algebra valued fermion $\rho$ is complex (i.e., not equal to $\tilde \rho$), unlike the previous case of $S_{YM,\Psi}$.  This  will change the physical content of the model. The gauge fixing of $\chi$ introduces just one set of ghosts $(\beta,\gamma)$, and we find the standard fixed currents for $S_{CS}$  and $S_\Psi$,
$$
u_t=\delta(\gamma)\rho\cdot t\, , \quad \tilde u_t=\delta(\gamma)\tilde\rho \cdot t \, , \quad u_ \epsilon=\delta(\gamma)\epsilon\cdot \Psi.
$$
The BRST descent leads to the following currents 
$$
 v_t=k\cdot\Psi \rho\cdot t + v^0_t\,,
\qquad \tilde v_t=k\cdot \Psi \tilde \rho\cdot t + \tilde v^0_t\, ,\quad V_\epsilon=\epsilon\cdot P+\epsilon\cdot\Psi k\cdot\Psi,
$$
where $v^0_t$ and $\tilde v^0_t $ denote the original $S_{CS}$ integrated vertex operators, so that
 $v_t$ and $\tilde v_t$ acquire a new term in $\Psi$. 

To impose GSO symmetry, we require invariance under flipping the sign of the  fields $\rho, \tilde{\rho}, q,y, \Psi, \chi $ and the corresponding ghosts.

Since we have untilded vertex operators $u_t,v_t$, and tilded ones $\tilde u_t, \tilde v_t$, the correlator will depend not only on the number of gluonic vertex operators versus gravity ones $u_\epsilon, v_\epsilon$, but also on the choice of whether the gluonic operators are of untilded or tilded type. Recall from the previous section that, for the theory $S_{CS}$ on its own, the only non-vanishing correlators were those with a single untilded operator and this led to a single comb colour structure that is equivalent to a single trace term.  This followed because of the need to have the same number of $\rho$s and $\tilde \rho$s in a nontrivial correlator and a single $\tilde \rho$ could only arise in one or both of the two fixed vertex operator.  Now single $\tilde\rho$s appear in $\tilde v_t$ and this essentially represents the coupling to gravity. Thus  the coupling to gravity introduces multiple trace terms, with the interaction between each single trace structure being mediated by gravity.  It is easy to see that with the $S_{CS,\Psi}$ system we can now have as many untilded vertex operators as we like with their number corresponding precisely to the number of traces.

\begin{thm}\label{cs-correlator}  Let the set $g$ index the gluons and $h$ the gravitons. To be non-vanishing, a $S_{CS,\Psi}$ correlator must have two fixed vertex operators, with the remaining ones integrated. The correlator of such a collection of vertex operators is  a sum over all partitions of the gluons into sets $T_1, T_2,\ldots, T_m$, where $m$ is the number of untilded gluonic vertex operators, and such that there is only one such vertex operator per $T_i$, $\cup_{i=1}^m T_i=g$, $|T_i|\geq 2$. Each allowed partition gives a contribution equal to \eqref{eqn:KPf}.
\end{thm}

Thus the correlator is the same as for $S_{YM,\Psi}$, except that there is a restriction on the allowed partitions of the gluons into traces.

\section{New Ambitwistor String Theories}
We can now assemble the full table of theories by combining the various possible choices of matter models on the left and right.  These can be identified with their  corresponding space-time theories by comparing the correlators to the formulae of CHY, and this results in table \ref{models}. Hopefully the acronyms for the models are self-explanatory except perhaps that BS denotes the bi-adjoint scalar $\phi^{aa'}$, where $a$ and $a'$ are respectively indices for the Lie algebras of $\SU(N) $ and $\SU(N')$, with action
$$
S_{\mathrm{BS}}= \int d^Dx \left( -\frac{1}{2}\p_\mu\phi^{aa'}\,\p^\mu\phi^{aa'} + \frac{1}{6}\, \phi^{aa'}\phi^{bb'}\phi^{cc'}f^{abc}f^{a'b'c'}\right)\, ,  
$$
where $f^{abc}$ and $f^{a'b'c'}$ are the structure constants of $\SU(N) $ and $\SU(N')$ respectively.

\begin{table}[h]{\small
\begin{tabular}{|c||l|l|l|l|l|}
  \hline
  \diagbox{$S^l$}{$S^r$}& $S_\Psi$ & $S_{\Psi_1,\Psi_2}$ & $S_{\rho,\Psi}^{(m')}$ & $S_{YM,\Psi}^{(N')}$ & $S_{YM}^{(N')}$\\ \hline \hline
  $S_\Psi$ & E &  & &  & \\ \hline 
  $S_{\Psi_1,\Psi_2}$ & BI & Galileon &  &  & \\ \hline  
  $S_{\rho,\Psi}^{(m)}$ & EM$\big|_{\text{U}(1)^{m}}$ & DBI & EMS$\big|_{\text{U}(1)^{m}\otimes \text{U}(1)^{m}}$ &  & \\ \hline 
  $S_{YM,\Psi}^{(N)}$ & EYM & extended DBI & EYMS$\big|_{\text{SU}(N)\otimes \text{U}(1)^{m'}}$ & EYMS$\big|_{\text{SU}(N)\otimes \text{SU}(N')}$ & \\ \hline  
  $S_{YM}^{(N)}$ & YM &  NLSM & YMS$\big|_{\text{SU}(N)\otimes \text{U}(1)^{m'}}$ & gen. YMS$\big|_{\text{SU}(N)\otimes \text{SU}(N')}$ & BS
  \\ \hline 
 \end{tabular}}
 \caption{Theories arising from the different choices of matter models.} \label{models}
\end{table}

Galileon theories are described by the action 
$$
S_{\mathrm{Galileon}}=\int d^Dx \left( -\frac{1}{2}\p_\mu\phi\,\p^\mu\phi + \sum_{m=3}^\infty g_m \phi \text{det}\{\p^{\mu_i}\p_{\nu_j}\phi\}_{i,j=1}^{m-1} \right),
$$
where $g_m$ are freely prescribable parameters.  However, our amplitudes only have one parameter. The theory that is singled out by our model is the one described in \cite{Cheung:2014dqa} in four dimensions, and in \cite{Cachazo:2014xea,Hinterbichler:2015pqa} in general dimension, which has smoother soft behaviour than the generic ones.

The Born-Infeld action is
$$
S_{\mathrm{BI}}=\int d^Dx\; \frac{1}{\ell^{2}}\left(\sqrt{-\text{det}(\eta_{\mu\nu}-\ell^2F_{\mu\nu})}-1\right)\,, $$ the Dirac Born-Infeld is
$$
S_{\mathrm{DBI}}=\int d^Dx \; \frac{1}{\ell^{2}} \left(\sqrt{-\text{det}\left(\eta_{\mu\nu}-\ell^2 \p_\mu\phi^a\p_\nu\phi^a -\ell F_{\mu\nu}\right)}-1\right)\, ,
$$
and the nonlinear-sigma model is  
$$
S_{\mathrm{NLSM}}=\int d^Dx\left( -\frac{1}{2}\text{tr}\left((\mathbb{1}-\lambda^2\Phi)^{-1}\p_\mu\Phi(\mathbb{1}-\lambda^2\Phi)^{-1}\p^{\mu}\Phi\right)\, \right), \qquad \mbox{
where } \Phi=\phi^at^a\, .
$$

In the table \ref{models}, we have only used $S_{YM}$.  Although this is sufficient to produce the correct tree-level amplitudes, it is an anomalous matter system and so has no hope to be extended beyond tree-level, and indeed its meaning as a string theory is unclear even at tree level.  We can obtain the same tree-amplitudes up to combinatorial factors by use of the comb system $S_{CS}$ and this is not anomalous.  However, this does lead to a doubling of the gauge degrees of freedom as described below in detail for the Einstein Yang-Mills system and bi-adjoint scalar.

 In table \ref{table-VO} we list the vertex operators in each model and the central charges.  Setting the central charge to zero gives the models that are critical and for which there is some reasonable hope that loop integrands can be described via these theories if they prove to be modular.

\subsection{Einstein Yang-Mills and $T^*$YM}

The worldsheet model that we discussed in the context of Einstein Yang-Mills theory, $S_{CS,\Psi}$, has a consistent quantisation. On the other hand, it does not correspond strictly to the building block of Einstein-Yang-Mills amplitudes, 
because only trace/comb structures consistent with the choice of untilded vertex operators are allowed. Attempts to find a theory that reproduces this correlator seem to lead back to the anomalous $S_{YM,\Psi}$ system. 

Since the theory $S_{CS,\Psi}$ presents no problems, and has correlators which match part of the Einstein-Yang-Mills building block, it is natural to ask whether it is related to a known theory. This theory must contain two types of gluons, associated to tilded and untilded vertex operators, and the untilded type must give the number of allowed multiple trace terms in an amplitude. These conditions are satisfied by the following spacetime action for the gauge field
\begin{equation}
\label{eq:tYM}
S_{T^*\text{YM}} = \int d^D x \;\tr( a_\mu\, D_\nu F^{\mu\nu}).
\end{equation}
The field $a_\mu$ is a Lagrange multiplier enforcing the Yang-Mills equations, $D_\nu F^{\mu\nu}=0$, and the action can be seen as a linearisation of the Yang-Mills action, $A_\mu\to A_\mu+a_\mu$. The field $A_\mu$ corresponds to the tilded degrees of freedom, and the field $a_\mu$ corresponds to the untilded ones. Since the propagator of this action connects $a_\mu$ to $A_\mu$ and the vertices contain a single $a_\mu$, the Feynman rules and a straightforward graph-theoretic argument show that there is one and only one $a_\mu$ external field per trace, also when the system is minimally coupled to gravity.

\subsection{Bi-adjoint scalar}

The use of the worldsheet system $S_{CS}$, with its two types of coloured currents, ${\tilde v}$ and $v$, is the reason for the Lagrange-multiplier-type action \eqref{eq:tYM}. An even simpler example is the bi-adjoint scalar theory, BS in table~\ref{models}. In this case, we can easily apply the procedure of \cite{Adamo:2014wea} and obtain explicitly the equations of motion. As in that paper, which was concerned with the Einstein theory, the spacetime background fields modify the worldsheet theory only through the constraints. The deformation of the constraints in the bi-adjoint scalar theory is particularly simple: the deformed ambitwistor constraint becomes
\begin{equation}
\mathcal{H} = P^2  \qquad \to \qquad \mathcal{H}_{(\phi,\Phi)}= P^2 
+ \Phi^{aa'}{\tilde v}^a{\tilde v}'^{a'}  + \phi^{aa'}{ v}^a{ v}'^{a'},
\end{equation}
where we introduced currents for each of the two independent groups SU($N$) and SU($N'$). The equations of motion are obtained as anomalies obstructing the vanishing of the constraint at the quantum level,
\begin{align}
\mathcal{H}_{(\phi,\Phi)}(\sigma)\,\mathcal{H}_{(\phi,\Phi)}(0) \sim \frac{1}{\sigma^2} 
\Big( & (2\,\partial^\mu\partial_\mu\Phi^{aa'} +f^{abc}f^{a'b'c'}\Phi^{bb'}\Phi^{cc'}) \, {\tilde v}^a {\tilde v}'^{a'} \nonumber \\
&+(2\,\partial^\mu\partial_\mu\phi^{aa'} +2\,f^{abc}f^{a'b'c'}\Phi^{bb'}\phi^{cc'}) \, v^a v'^{a'} \Big)(0) \nonumber \\
+ & \;\;\text{simple} \;\; \text{pole}.
 \end{align}
If the equations of motion hold, there is no double pole and in fact the OPE is finite, because there can be no simple pole in the self-OPE of a bosonic operator in the absence of higher poles.
The spacetime action associated to these equations of motion takes the Lagrange-multiplier form
\begin{equation}
S_{\text{BS}} = \int d^D x \; \phi^{aa'} \left(\partial^\mu\partial_\mu\Phi^{aa'} +\frac{1}{2}\, f^{abc}f^{a'b'c'}\Phi^{bb'}\Phi^{cc'}\right) .
\end{equation}
It should be seen as the analogue of the gauge theory action \eqref{eq:tYM}.

\begin{table}[h]  {\small
\begin{tabular}{|l||l|l|}\hline
 Theories & Integrated vertex operators & Central charge $c$  \\ \hline\hline
 E &  $V_h=\left(\epsilon\cdot P + k\cdot\Psi\epsilon\cdot\Psi\right)\left(\tilde\epsilon\cdot P + k\cdot\tilde\Psi\tilde\epsilon\cdot\tilde\Psi\right)$ & $3(d-10)$  \\ \hline
 \multirow{2}{*}{EM} & $V_h,\, V_\gamma$ & \multirow{2} {*}
 {$3(d-10+\frac{m}{6})$}  \\
 & $V_\gamma=\left(k\cdot \Psi\,t\cdot \rho \right) \left(\tilde\epsilon\cdot P + k\cdot\tilde\Psi \tilde\epsilon\cdot\tilde\Psi\right)$ & \\ \hline
 \multirow{2}{*}{EMS} & $V_h,\,V_\gamma,\,V_{\tilde\gamma},\,V_S$ & \multirow{2}{*}{$3(d-10+\frac{m+\tilde{m}}{6})$} \\ 
 & $V_S=\left(k\cdot \Psi \,t\cdot \rho\right)\left(k\cdot \tilde\Psi \tilde\rho\cdot t\right)$  &\\ \hline
 BI & $V_{BI}=\left(k\cdot \Psi_1 k\cdot \Psi_2\right)\left(\tilde\epsilon\cdot P + k\cdot\tilde\Psi \tilde\epsilon\cdot\tilde\Psi\right)$ & $\frac{1}{2}\left(7d-38\right)$ \\ \hline
 Galileon & $V_G=\left(k\cdot \Psi_1 k\cdot \Psi_2\right)\left(k\cdot \tilde\Psi_1 k\cdot \tilde\Psi_2\right)$ & $4d-8$ \\ \hline
 \multirow{2}{*}{DBI} & $V_{BI},\,V_{S_{BI}}$ & \multirow{2}{*}{$\frac{1}{2}(7d+m-38)$} \\
  & $V_{S_{BI}}=\left(k\cdot \Psi_1 k\cdot \Psi_2\right)\left(k\cdot \tilde\Psi\,t\cdot \tilde\rho\right)$ & \\ \hline
 \multirow{2}{*}{$T^*$YM}&$V_g=\left(\half t\cdot [\rho,\rho]\right)\left(\tilde\epsilon\cdot P + k\cdot\tilde\Psi\tilde\epsilon\cdot\tilde\Psi\right)$ & \multirow{2}{*}{$\frac{5}{2}(d-12)$} \\ 
 & $V_{\tilde{g}}=\left(t\cdot\left( [\rho,\tilde\rho]+ [q,y]\right)\right)\left(\tilde\epsilon\cdot P + k\cdot\tilde\Psi\tilde\epsilon\cdot\tilde\Psi\right)$ & \\ \hline
 \multirow{3}{*}{E$T^*$YM} & $V_h,\,V_g,\,V_{\tilde{g}}$ & \multirow{3}{*}{$3(d-10)$}  \\
 &$V_g=\left(k\cdot\Psi\,t\cdot \rho +\half t\cdot [\rho,\rho]\right)\left(\tilde\epsilon\cdot P + k\cdot\tilde\Psi\tilde\epsilon\cdot\tilde\Psi\right)$ &\\ 
 & $V_{\tilde{g}}=\left(k\cdot\Psi \,t\cdot \tilde\rho +t\cdot\left( [\rho,\tilde\rho]+ [q,y]\right)\right)\left(\tilde\epsilon\cdot P + k\cdot\tilde\Psi\tilde\epsilon\cdot\tilde\Psi\right)$ & \\ \hline
 \multirow{2}{*}{NLSM} & $V=\left(\half t\cdot [\rho,\rho]\right)\left(k\cdot \tilde\Psi_1 k\cdot \tilde\Psi_2\right)$ & \multirow{2}{*}{$3d-19$}  \\
 & $\tilde{V}=\left(t\cdot\left( [\rho,\tilde\rho]+ [q,y]\right)\right)\left(k\cdot \tilde\Psi_1 k\cdot \tilde\Psi_2\right)$ &\\\hline
\end{tabular}}
\caption{Table of the different theories and their integrated vertex operators.}
\label{table-VO}
\end{table}

\section{Discussion}
There are many issues to explore further and we briefly mention a few of them here.  We have listed the central charges for the various models that are not already anomalous in table \ref{table-VO}.  It can be seen that many of the models have some critical dimension where the central charge vanishes.  Indeed, one can often simply add some number of Maxwell fields to make them critical if one starts in low enough dimension.  This suggests that a number of these models might give rise to plausible string expressions for corresponding loop integrands such as given in \cite{Adamo:2013tsa} for the type II theory in 10 dimensions.  However, an independent criterion is that the loop integrand so obtained should be modular invariant and this may well exclude many of the critical models as it does in conventional string theory.  

There is also the question as to whether there are further vertex operators that we have missed and therefore further sectors of these theories.  For the 10 dimensional models, following \cite{Adamo:2013tsa}, one can introduce a spin field $\Theta^\alpha$ associated to each $\Psi$ field and use these to introduce further vertex operators that will correspond to space-time fields with spinor indices.  For the type II Einstein theory these give rise to the Ramond sector vertex operators \cite{Adamo:2013tsa} and it can be seen that the same procedure can be applied more generally to some of the models here, particularly the  Einstein $T^*$YM models.  Following the same procedure one then extends the Einstein NS sector to include the Ramond sectors of type II gravity theories.  However, we can see that the $T^*$YM vertex operators can only be extended in this way on the one side corresponding to the spin operator constructed from the $\Psi$ in the Yang-Mills vertex operator.  Thus one supersymmetry acts trivially on the Yang-Mills and hence is degenerate (it does not square to provide the Hamiltonian on the Yang-Mills fields). 

By extending the worldsheet matter fields we have generated new possible couplings to space-time fields.  It would be interesting  to explore whether these couplings can be made consistent in the fully nonlinear regime as described in \cite{Adamo:2014wea,Chandia:2015sfa}.

There remain other formulae based on the scattering equations, for which  an underlying ambitwistor string theory has not yet been found. It would for example be  interesting to find ambitwistor strings that give rise to the class of formulae with massive legs \cite{Dolan:2013isa,Naculich:2014naa,Naculich:2015zha, Naculich:2015coa}, and that for ABJM theory \cite{Huang:2012vt,Cachazo:2013iaa}, although see the twistor string \cite{Engelund:2014sqa}.

Perhaps the most irritating issue is that we have not been able to find an Einstein-Yang-Mills model that is anomaly-free without unwanted linearized modes.  Conventional string theory produces such amplitudes in open string theory and in closed string heterotic models.  However, the ambitwistor heterotic string has corrupt gravity amplitudes and so far there has been no ambitwistor analogue of open strings. Nevertheless the $T^*$YM model is likely to make sense and provide the correct amplitudes at 1-loop if modular, although the pure gauge sector does not have loop amplitudes beyond 1-loop.

\subsection{Acknowledgements}
We are grateful to David Skinner and Ellis Yuan for many contributions and to Song He and Jaroslav Trnka for useful discussions.  EC is supported in part by the Cambridge  Commonwealth, European and International Trust, YG is supported by the EPSRC and the Mathematical Prizes fund, LM is partially supported by EPSRC grant number EP/J019518/1, and RM is supported by a Marie Curie Fellowship and a JRF at Linacre College.  KR is supported in part by a Marie Curie Career Integration Grant (FP/2007-2013/631289).

\appendix

\section{Correlators for \texorpdfstring{$S_{YM,\Psi}$}{S-YM,Psi}}\label{sec:ecomb-proof}
Here we give the proof of  theorem \ref{ym-correlator}. In the main text several versions of the present idea are realized. We will demonstrate and prove the mechanism in the simplest setting, which already contains all necessary ingredients, and comment on adaptations and restrictions afterwards. Concretely we use a single free fermion $\rho^a $ and a generic level zero current $j^a$. The fields have the same OPEs as above, that is $j^a$ form a current algebra and $\rho^a$ are in the adjoint presentation of the $j$-algebra, i.e.
\begin{equation}\label{eqn:appendixA-OPEs}
\rho^a(\sigma) \rho^b(0) \sim \frac{1}{\sigma} \delta^{ab}  ~, \qquad j^a(\sigma) j^b(0) \sim \frac{1}{\sigma} f^{abc} j^c  ~, \qquad j^a(\sigma) \rho^b(0) \sim \frac{1}{\sigma} f^{abc} \rho^c  ~.
\end{equation}

The strategy of the proof is as follows: both the full space-time amplitude $\mathcal{A}(g,h)$ and the world-sheet correlator $A(g,h)$ are a (multiple) sum of simple terms. The sum in $\mathcal{A}$ is over trace sectors as well as a choice of gluon labels, while the sum in $A(g,h)$ is simply the Wick expansion of the expectation value. Schematically we get
\begin{equation}
\mathcal{A} = \sum_{x \in X} \mathcal{A}(x) \qquad \qquad \text{and} \qquad \qquad A = \sum_{y\in Y} A(y)
\end{equation}
where $X,Y$ are sets labelling the trace sectors  and organization of sets of Wick contractions respectively. 
Then we will show that $x \in X \Rightarrow x \in Y$ and $y \in Y \Rightarrow y \in X$. Upon showing that each element in $X,Y$ is unique we get $X=Y$. Along the way we will see that $A(x) = \mathcal{A}(x)$, hence establishing $A = \mathcal{A}$.

To clarify the structure of the discussion we firstly only insert integrated vertex operators on the world-sheet -- which corresponds to considering the full Pfaffian in the CHY formula -- keeping in mind that to get a non-vanishing result we need to go over to the reduced Pfaffian. That step will be taken at the end.

So we will have to examine the correlation function of two types of operators,
\begin{equation}\label{eqn:appendixA-one}
\mathcal{O}^{gl} = k \cdot \Psi ~ t \cdot \rho + t \cdot j \qquad \text{and} \qquad  \mathcal{O}^{gr} = k \cdot \Psi ~ \epsilon \cdot \Psi + \epsilon \cdot P  ~,
\end{equation}
for (one half of) the gluon and graviton integrated vertex operators respectively. The claim is that the string-worldsheet correlator
\begin{equation}\label{eqn:appendixA-two}
A( g , h) := \left\langle ~ \prod_{a \in g} \mathcal{O}^{gl}_a  ~ ~ \prod_{a \in h}  \mathcal{O}^{gr}_a  ~ \right\rangle
\end{equation}
where $g$ and $h$ are the sets containing the gluon and graviton labels respectively, is equal to (one part of the CHY representation of) the full space-time amplitude
\begin{equation}\label{eqn:appendixA-three}
\mathcal{A} = \sum_{\text{trace sectors}} \mathcal{C}_1 \cdots \mathcal{C}_m ~ \pf \Pi ~,
\end{equation}
where the sum goes over all trace sectors possible. In particular, it includes a sum over the number of traces $m = 1, \cdots , [ |g| /2 ]$. The matrix $\Pi$, defined in \cite{Cachazo:2014xea}, of course depends on the trace sector.

The main step in going between the representations two is the identity \cref{eqn:Comb-Trace-relation}, which we repeat here for the readers convenience
\begin{equation}\label{eqn:appendixA-four}
\sigma_{ab} ~ \mathcal{C} (T) = \mathcal{K}( b,a | T) ~,
\end{equation}
with $\mathcal{K}$, the `comb structure' defined in the main text. Its arguments are the unordered set $T$ and two of its elements, $a,b \in T$. Using the anti-symmetry and multi-linearity of the Pfaffian, expression \cref{eqn:appendixA-three} can be brought into the form
\begin{equation}\label{eqn:appendixA-five}
\sum_{\substack{\text{trace}\\ \text{sectors}}}  \sum_{\substack{a_1 < b_1 \in T_1 \\ \cdots \\ a_m < b_m \in T_m }}  \mathcal{K}(a_1 ,b_1 | T_1) \cdots  \mathcal{K}(a_m ,b_m | T_m)  ~~ \pf M( h , \{a_i \} , \{b_i\} | h ) ~.
\end{equation}
This is the representation of the amplitude which the world-sheet correlator \cref{eqn:appendixA-two} will land us on.

Let us now consider evaluating the correlator $A(g,h)$. We will see that it gives rise to a multiple sum over terms, which turn out to be the same that \cref{eqn:appendixA-five} sums over. The first step is to expand the product of all the $\mathcal{O}^{gl}$s into a sum. The sum is over all ways of putting either a $k\Psi \rho$ or a $j$ at each gluon insertion. This is a binary choice so it leads to $2^{|g|}$ terms. Name the set of gluon labels which carry a $k\Psi \rho$ insertion by $e$ for each term. The path integral over the $\Psi $ field can now be performed for each term individually. Since $\Psi$ is fermionic, the path integral vanishes unless $|e|$ is even. Define $m := |e| /2$, which is integer. The result of this path integral is of course simply a factor of
\begin{equation}\label{eqn:appendixA-six}
\pf M( h , e | h ) 
\end{equation}
for each term in the sum, by the standard reasoning described for example in \cite{Mason:2013sva}. note that, since $e$ only appears once, the Pfaffian depends on the ordering of the elements in $e$. Now the correlator $A$ is a sum over ways of partitioning $g$ into $e$ and $g-e$, with the condition that $|e|$ be even, and each term in the sum looks like\footnote{From now onwards we omit the colour structure and abbreviate $t_a \cdot \rho (\sigma_a) = \rho_a$ and $t_a \cdot j (\sigma_a) = j_a$.}
\begin{equation}\label{eqn:appendixA-seven}
\langle ~ \prod _{a \in e} \rho_a  ~ \prod_{a \in g-e} j_a ~ \rangle ~~ \pf M( h , e | h ) ~.
\end{equation}
It should be clear that the remaining worldsheet correlator will give rise the product of $\mathcal{K}$s and the remaining sum. Let us see how this happens in detail. Performing the Wick expansion of the $\rho,j$ correlator breaks it down into a product of smaller pieces, so far until each factor contains precisely two (i.e. a pair of) $\rho$ insertions accompanied by some subset of the $j$ insertions. Label the pair of $\rho$ insertions int the $i^{th}$ factor by $a_i,b_i$ and the accompanying set of $j$ insertions by $T_i$. Wick expansion makes sure that each choice of pairs and each choice of accompanying $j$ insertions appears at least once and only once. Schematically we get
\begin{equation}\label{eqn:appendixA-eight}
\langle ~ \prod _{a \in e} \rho_a  ~ \prod_{a \in g-e} j_a ~ \rangle  = \sum_{\text{partitions}} ~ \prod_{i=1}^m ~ \langle ~ \rho_{a_i} \, \rho_{b_i} ~ \prod_{c_i \in T_i} j_{c_i} ~\rangle   ~.
\end{equation}
The remaining correlator is now easily evaluated using the OPEs \cref{eqn:appendixA-OPEs} to give
\begin{equation}\label{eqn:appendixA-nine}
 \langle ~ \rho_{a} \, \rho_{b} ~ \prod_{\substack{c \in T \\ c \neq a,b }} j_{c} ~\rangle  =  \mathcal{K}(a , b | T) ~.
\end{equation}
Note that the symmetry properties of the function $\mathcal{K}$ in its arguments naturally arise from the statistics of the fields $\rho,j$.

Actually, performing the Wick expansion in \cref{eqn:appendixA-eight} does not preserve the order of the $\rho$ insertions, so, as they are fermions, a factor of $(-1)$ might appear. We can absorb this factor by bringing the rows/columns of the matrix $M$ into the same order as the $\rho$ appear on the \emph{rhs} of \cref{eqn:appendixA-eight}. Then \cref{eqn:appendixA-seven} becomes
\begin{equation}
\sum ~  \mathcal{K}(a_1 , b_1 | T_1) \cdots \mathcal{K}(a_m , b_m | T_m) ~~ \pf M( h , \{ a_1 ,b_1, \cdots , a_m,b_m\} | h) ~,
\end{equation}
which is precisely the summand appearing in the full space-time amplitude\footnote{In fact there will be additional sign factors from permutations the rows/columns in the Pfaffian.}. We repeat that Wick expansion ensures that every possible configuration of the summand is summed over, each term appearing at least once and only once. 


We have shown that the expressions $A$ and $\mathcal{A}$ are sums over the same simple terms, involving $\mathcal{K}$s and the corresponding $\pf M$. To clarify, on one hand, the sum in $A$ goes over different ways of choosing $m$ pairs $a_i,b_i$ from $g$ and different ways of forming $m$ unordered sets $T_i$ from the labels left over, as well as the sum over $m$. The set $X$ mentioned above contains as elements the ways of making such choices. On the other hand, the sum in $\mathcal{A}$ goes over ways of splitting the labels $g$ into $m$ unordered subsets $T_i$ and the ways of picking a pair from each subset, as well as the sum over $m$. The set of these choices is $Y$.  What remains to show is that the sums are actually the same or, equivalently, that each sum includes the other. To do so, we go back to the full expressions
\begin{equation}
A( g , h) = \left\langle ~ \prod_{a \in g} \mathcal{O}^{gl}_a  ~ ~ \prod_{a \in h}  \mathcal{O}^{gr}_a  ~ \right\rangle  \qquad \text{and} \qquad \mathcal{A}(g,h) = \sum_{\text{trace sectors}} \mathcal{C}_1 \cdots \mathcal{C}_m ~ \pf \Pi ~.
\end{equation}
and argue that if a term appears in $A$ it also appears in $\mathcal{A}$ and vice versa. 
Additionally we argue that each term appears at least once and only once in each expression, which will conclude the proof that they are equal.

It is clear that both sums contain the summation over $m=1, \cdots , [ |g|/2 ]$ in them, which is to be understood as the number of traces. Take a contribution from $A$ with $|e| =2 m$. Each term in the sum is uniquely determined my specifying $m$ pairs $\{a_i,b_i\}$ and $m$ unordered sets $T_i$. As mentioned previously, Wick expansion guarantees that each term appears once and only once. A given term should have a partner in $\mathcal{A}$ at $m$ traces. Looking at the representation \cref{eqn:appendixA-three} this is not straightforward to see, but the equivalent representation \cref{eqn:appendixA-five} makes this readily apparent. The sum over trace sectors will include one term where the $T_i$ in $\mathcal{A}$ are precisely\footnote{In a slight abuse of notation, what is called $T_i$ in $\mathcal{A}$ is actually $T_i \cup \{ a_i, b_i\} $ in $A$.} the $T_i$ in $A$ whereupon the sums $a_i,b_i \in T_i$ will contain one term in which all $a_i,b_i$ in $\mathcal{A}$ agree with those in $A$. This shows that each term in $A$ has a partner in $\mathcal{A}$, establishing the statement $ y \in Y \Rightarrow y \in X$. Of course Wick expansion guarantees the uniqueness of the terms in $Y$.

Conversely, one term in the summation in $\mathcal{A}$ is uniquely specified by fixing a trace structure and picking out one term of the summations over $a_i,b_i$. In other words, it is specified by a collection of $m$ sets $T_i$ and a choice of pairs $\{a_i,b_i \}$ for each set. To see that any such term is also contained in $A$ simply notice that the above data uniquely specifies a term in $A$ via
\begin{equation}
\prod_{i=1}^m \langle ~ \rho_{a_i} \rho_{b_i} ~ \prod_{c_i \in T_i} j_{c_i} ~ \rangle ~~ \pf M(h , \{ \{a_i,b_i \} _i \} | h) ~.
\end{equation}
Hence, each term in $\mathcal{A}$ has a partner in $A$ and this establishes the statement $ x \in X \Rightarrow x \in Y$. The uniqueness of each element follows by construction.

\subsection{The Reduced Pfaffian}

 The Pfaffian we discussed so far actually vanishes for physical systems, i.e. when momentum conservation, gauge invariance and the scattering equations hold. Hence it is replaced by the reduced Pfaffian  $\pfr  \Pi$ defined in either of the following equivalent ways
 \begin{equation}
\label{eq:redpf}
 \pfr \Pi := \pf \Pi_{i,j^\prime} = \frac{(-)^a}{\sigma_a} \pf \Pi_{a,i}  = - \frac{(-)^a}{\sigma_a} \pf \Pi_{a,j^\prime} = \frac{(-)^{a+b}}{\sigma_{ab}} \pf \Pi_{a,b}
 \end{equation}
where $a,b$ label gravitons, with the restriction to not remove any row/column of the matrix $B$, and the $i, j^\prime$ label traces. On the other hand, we know that the expectation value of all integrated vertex operators will also vanish, and we have to insert precisely two fixed vertex operators. For an all graviton amplitude, this was discussed in \cite{Mason:2013sva}. It follows from BRST invariance that the amplitude is invariant under the choice of which vertex operators to take fixed/integrated. Hence, if there are at least two gravitons and arbitrarily many gluons, the full amplitude must be equal to the CHY formula. We will now show that the reduced Pfaffian also follows when using fixed vertex operators for two gluons or one gluon and one graviton, trying to present the following expressions in a suggestive form.

\subsubsection{Two Gluons Fixed}
Denote the labels of the fixed gluon operators as $c,d$. With the reduced Pfaffian defined as
\begin{equation}
\pfr \Pi = \pf \Pi_{i,j^\prime}
\end{equation}
there are two cases, $j^\prime = i$ or $j^\prime \neq i$. In the first case the trace $T_i$ is totally removed from the Pfaffian and we can write
\begin{equation}
\cdots \, \mathcal{C}_i \, \cdots ~ \pf \Pi_{i,i^\prime} = \frac{1}{(d \, c)} \, \cdots  \,  \mathcal{K}(c,d | T_i) ~ \pf \Pi_{i,i^\prime}  ~,
\end{equation}
with the gluons $c,d$ being members of the trace $T_i$. The factor $\frac{1}{(d \, c)}$ fits into the interpretation of \cite{Mason:2013sva} as ghost field correlator. Note that there is no sum over choices of pairs in $T_i$, instead the comb $\mathcal{K}$ appears with fixed start/end points, corresponding to the insertion of fixed vertex operators for the gluons $c,d$.

In the second case ($j^\prime \neq i$), name the traces such that $c \in T_1$ and $d \in T_2$. Now each term in the expansion of the worldsheet correlator will look like (omitting all irrelevant factors)
\begin{equation}
\begin{aligned}
  \frac{1}{\sigma_{cd}} \sum_{\substack{a\in T_1 \\ b \in T_2}} & \mathcal{K}(c,a | T_1)      \mathcal{K}(d,b | T_2) ~ \pf ( a , b , \cdots  )   =  \mathcal{C}(T_1) \, \mathcal{C}(T_2) \, \sum_{\substack{a\in T_1 \\ b \in T_2}}  \frac{\sigma_{ac} \sigma_{bd} }{\sigma_{cd} } ~ \pf ( a , b , \cdots  ) \\
  &    =  \mathcal{C}(T_1) \, \mathcal{C}(T_2) \, \sum_{a \in T_1 }  \frac{\sigma_{ac} }{\sigma_{cd} } ~ \pf ( a , \sum _{b \in T_2 } \sigma_{bd} \,  b , \cdots  )    \\ 
  &  =  \mathcal{C}(T_1) \, \mathcal{C}(T_2) \, \sum_{a \in T_1 }  \frac{\sigma_{ac} }{\sigma_{cd} } ~ \pf ( a , - \sum _{b \in T_1 } \sigma_{bd} \,  b , \cdots  )  \\  
& =  \mathcal{C}(T_1) \, \mathcal{C}(T_2) \, \sum_{\substack{a\in T_1 \\ b \in T_1}}  \frac{\sigma_{ac} \sigma_{db} }{\sigma_{cd} } ~ \pf ( a , b , \cdots  )   \\ 
&  =  \mathcal{C}(T_1) \, \mathcal{C}(T_2) \, \sum_{ a< b \in T_1 }  \frac{\sigma_{ac} \sigma_{db} - \sigma_{bc} \sigma_{da} }{\sigma_{cd} } ~ \pf ( a , b , \cdots  )    \\
& =  \mathcal{C}(T_1) \, \mathcal{C}(T_2) \, \sum_{ a< b \in T_1 }  \sigma_{ba}  ~ \pf ( a , b , \cdots  )    \equiv  \mathcal{C}(T_1) \, \mathcal{C}(T_2) ~ \pf \Pi_{2,2^\prime} ~.
\end{aligned}
\end{equation}
Note that we had to use the scattering equations and the antisymmetry of the Pfaffian to arrive at the final result.

\subsubsection{One Gluon, One Graviton Fixed}
The computation for fixing one gluon and one graviton vertex operator is largely analogous to the previous one. Moreover, BRST invariance guarantees that the final result will be as desired. Let us nevertheless demonstrate the necessary manipulations. Denote the fixed gluon by $c$, with $c \in T_1$, and the fixed graviton by $m$
\begin{equation}
\begin{aligned}
 \frac{1}{\sigma_{mc}} \sum_{a\in T_1} \mathcal{K}(a,c |T_1) &~ \pf ( a , \cdots , \check{m} , \cdots )  = \mathcal{C}(T_1) \sum_{a\in T_1 } \frac{\sigma_{ca}}{\sigma_{mc}} ~ \pf ( a , \cdots , \check{m} , \cdots ) \\
& = \mathcal{C}(T_1) \frac{1}{\sigma_{mc}}   ~ \pf (\sum_{a\in T_1 }  \sigma_{ca} \, a , \cdots , \check{m} , \cdots ) \\
&  = \mathcal{C}(T_1) \frac{1}{\sigma_{mc}}   ~ \pf ( - \sigma_{cm} \, m , \cdots , \check{m} , \cdots )   \\
& = \mathcal{C}(T_1)   ~ \pf (   m , \cdots , \check{m} , \cdots )   = \mathcal{C}(T_1)   ~ \pf \Pi _{1,1^\prime} ~.
\end{aligned}
\end{equation}
Again we had to make use of the scattering equations.

\subsection{Adaption and Restriction}
As mentioned in the text, it seems not to be possible to find a level zero current via descent from $\rho $ in a consistent way satisfying \cref{eqn:appendixA-nine}. Hence, the main text contains an adaption of the system discussed above, using two fermions $\rho^a , \tilde{\rho}^a$, conjugate to each other. Via the descent, $\rho^a$ gives rise to $j^a$ while $\tilde{\rho}^a$ gives rise to $\tilde{j}^a$. The OPEs between the currents and the fields are
\begin{equation}
\begin{aligned}
& \rho^a(z) j^b(0) \sim \frac{1}{z} f^{abc} \rho^c ~, ~~ \tilde{\rho}^a(z) j^b(0) \sim \frac{1}{z} f^{abc} \tilde{\rho}^c ~,~~  \\
&\rho^a(z) \tilde{j}^b(0) \sim \frac{1}{z} f^{abc} \tilde{\rho}^c ~, ~~ \tilde{\rho}^a(z) \tilde{j}^c(0) \sim 0 ~.
\end{aligned}
\end{equation}
We shall now examine the correlators of this system.

\begin{figure}[h]
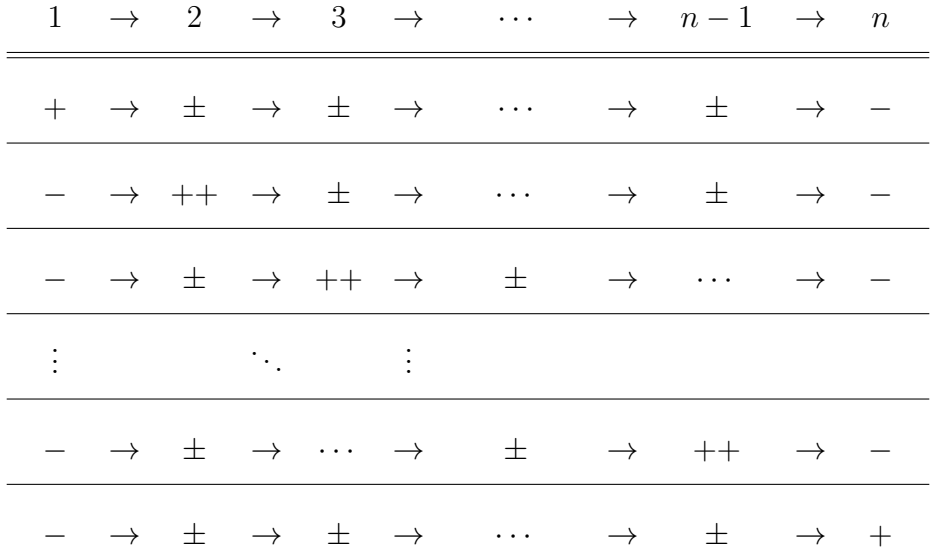

\begin{equation*}
\begin{array}{ccccccccccc}
\quad  1  \quad &  \to &  \quad  2 \quad & \to &   \quad    3 \quad & \to &   \qquad    \cdots \qquad & \to &  \quad  n-1 \quad & \to &    \quad n  \quad \\  \hline\hline 
\quad  + \quad &  \to & \quad  \pm \quad &  \to &  \quad    \pm \quad &  \to &  \quad    \cdots \quad & \to &  \quad  \pm \quad &  \to &   \quad -  \quad \\  \hline
- &  \to & ++ & \to &  \pm & \to &  \cdots & \to &  \pm & \to &  - \\  \hline
- &  \to & \pm & \to &  ++ & \to &  \pm & \to &  \cdots & \to &  - \\  \hline
\vdots & • & • & \ddots & • & \vdots \\  \hline
- & \to &  \pm & \to &  \cdots & \to &  \pm & \to & ++ & \to &  - \\  \hline
- & \to &  \pm & \to &  \pm & \to &  \cdots &  \to & \pm &  \to & + 
\end{array} 
\end{equation*}
\caption{Charge flows} \label{charge-flows}
\end{figure}

First, note that by taking the fixed vertex operators to be $(\rho + \tilde{\rho} )$, the discussion above would carry over verbatim. There is a crucial difference however. The current appearing in the associated integrated vertex operator does not quite satisfy \cref{eqn:appendixA-nine}, but instead
\begin{equation}
 \langle ~ (\rho_{a} + \tilde{\rho}_a ) \,(\rho_{b} + \tilde{\rho}_b )  ~ \prod_{c \in T} ( j_{c} + \tilde{j}_c) ~\rangle  = ~ |T| \,  ~ \mathcal{K}(a , b | T) ~.
\end{equation}
So each contribution from a different trace sector will come with a different prefactor $\prod_i^m |T_i|$, spoiling the relative coefficient between partial amplitudes. As the prefactor depends on the given partition of particles into traces, it cannot be removed by a field rescaling. The origin of this factor can be understood by simply counting the ways in which a full comb can be generated. If we represent the fields $\rho,\tilde{\rho}$ by $+,-$ and the currents $j,\tilde{j}$ by $\pm, ++$ respectively, the possible contractions can be found by drawing all allowed charge flows as in figure \ref{charge-flows}

Observe that each contraction must have exactly one insertion of $\tilde{j}$ (represented by $++$) or $\tilde{\rho}$ (represented by $+$), independent of the length $n$ of the chain, while there are $n-1$ insertions of $j$ or $\rho$. Summing over the possible positions of the tilded operator in the chain gives rise to the over-counting by $|T|$. Note that each contraction contributes exactly the same analytical \& colour structure.

Having understood the (non--trivial) origin of the factor $|T|$, the remainder of the discussion, showing how to remove it, follows trivially. Denote $v$ the vertex operator containing $\rho$ and $j$ and  $\tilde{v} $ the one containing $\tilde{\rho}$ and $\tilde{f}$, either integrated or fixed. It is now clear that choosing to insert $\tilde{v}$ at $m$ of the gluon punctures and $v$ at the others will give rise (following the general discussion above) to the complete color ordered partial amplitude with $m$ traces
\begin{equation}
\mathcal{C}(T_1) \cdots \mathcal{C}(T_m) ~ \pfr \Pi ~,
\end{equation}
which concludes the discussion.

\bibliography{twistor-bib}  
\bibliographystyle{JHEP}
\end{document}